\documentclass[amssymb,amsmath,showpacs,superscriptaddress,preprint]{revtex4-1}
\usepackage{graphicx}
\usepackage{amsmath}
\usepackage{gensymb}
\usepackage{xcolor}
\usepackage{blindtext}
\usepackage[allcolors=blue]{hyperref}%
\newcommand{\BSS}[1]{{\color{black}#1}} 

\begin{document}

\title{
Mechanical Amorphization of Glass-Forming Systems Induced by Oscillatory Deformation: The Energy Absorption and Efficiency Control
}

\author{Baoshuang Shang$^\#$}
\email{shangbaoshuang@sslab.org.cn} 
\affiliation{Songshan Lake Materials Laboratory, Dongguan 523808, China}
\author{Xinxin Li$^\#$}
\email{xinxinli@hku.hk}
\affiliation{Songshan Lake Materials Laboratory, Dongguan 523808, China}
\affiliation{Department of Mechanical Engineering, The University of Hong Kong, Pokfulam Road, Hong Kong SAR, China}
\author{Pengfei Guan}
\email{pguan@nimte.ac.cn}
\affiliation{Advanced Interdisciplinary Science Research Center (AiRCenter) , Ningbo Institute of Materials Technology and Engineering, Chinese Academy of Sciences, Ningbo 315201, China}
\affiliation{Beijing Computational Science Research Center, Beijing 100193, China}
\author{Weihua Wang}
\affiliation{Songshan Lake Materials Laboratory, Dongguan 523808, China}
\affiliation{Institute of Physics, Chinese Academy of Sciences, Beijing 100190, China}
\date{\today}
\begin{abstract}
The kinetic process of mechanical amorphization plays a central role in tailoring material properties. Therefore, a quantitative understanding of how this process depends on loading parameters is critical for optimizing mechanical amorphization and tuning material performance. In this study, we employ molecular dynamics simulations to investigate oscillatory deformation-induced amorphization in three glass-forming intermetallic systems, addressing two unresolved challenges: (1) the relationship between amorphization efficiency and mechanical loading, and (2) energy absorption dynamics during crystal-to-amorphous (CTA) transitions. Our results demonstrate a decoupling between amorphization efficiency—governed by work rate and described by an effective temperature model—and energy absorption, which adheres to the Herschel–Bulkley constitutive relation. Crucially, the melting enthalpy emerges as a key determinant of the energy barrier, establishing a thermodynamic analogy between mechanical amorphization and thermally induced melting. This relationship provides a universally applicable metric to quantify amorphization kinetics. By unifying material properties, and loading conditions, this work establishes a predictive framework for controlling amorphization processes. These findings advance the fundamental understanding of deformation-driven phase transitions and offer practical guidelines for designing materials with tailored properties for ultrafast fabrication, ball milling, and advanced mechanical processing techniques.
\end{abstract}
\maketitle
\def\thefootnote{\#}\footnotetext{These authors contributed equally to this work}\def\thefootnote{\arabic{footnote}}
\section{Introduction}
 Mechanical amorphization is an ubiquitous phenomenon during the material fabrication process, which goes from the crystal state to the amorphous state with the help of external loading\cite{angell1995formation,Suryanarayana2001,Idrissi2022on,Li2022Amorphization}. This process generates hierarchical structures characterized by a gradient in grain size, spanning from the micrometer to nanometer scale, and may ultimately yield entirely amorphous configurations\cite{Chen2003shock,Zhao2018ShockinducedAI,Zhao2015pressure,Li2023ultrafast}. In particular, mechanical amorphization can also facilitate the synthesis of new phase states\cite{Tang2021synthesis,SanMiguel2021,RosuFinsen2023medium,Luo2023polyamorphism}.
 The underlying mechanisms are generally attributed to high-density dislocation jamming \cite{He2016Insitu,Zhao2017generating} or mechanical instability \cite{Wang2016Insitu,Bu2024Elastic}.

Recent advancements reveal unexpected behaviors in specific intermetallic systems. Luo et al. \cite{Luo2019plasticity} demonstrated that mechanical amorphization can proceed in the absence of dislocations. Subsequent research by Hu et al. \cite{Hu2023amorphous} further established that this phenomenon can serve as a plasticity mechanism in specific glass-forming systems. Additionally, Li et al. \cite{Li2024stress} demonstrated that the correlation between mechanical amorphization and glass-forming ability can be modulated by external stress. These findings highlight the sensitivity of mechanical amorphization to both intrinsic material properties and extrinsic loading conditions. Consequently, elucidating this process in glass-forming systems may provide critical insights to achieve the ductility of brittle materials \cite{Zhao2023amorphization}.
Recent investigations of oscillatory deformation-induced amorphization \cite{Li2023ultrafast,Li2024stress} have provided new insights. First, the crystal-to-amorphous (CTA) transition becomes precisely controllable under cyclic loading, with the degree of amorphization increasing monotonically with the number of cycles. Second, mild oscillatory deformation minimizes thermal effects, enabling isolation of the mechanical contribution. However, how the mechanical amorphization evolves with loading parameters of oscillatory deformation and the mechanism of mechanical amorphization induced by oscillatory deformation is still elusive. 

To address the aforementioned challenges and systematically investigate (1) the relationship between amorphization efficiency and mechanical loading, and (2) energy absorption dynamics during CTA transitions, we employ molecular dynamics simulations to study oscillatory deformation-induced amorphization in three intermetallic systems with high simulated glass-forming ability. Our results demonstrate a decoupling between the amorphization efficiency and energy absorption with respect to the external loading parameters. Specifically, amorphization efficiency is governed by the applied work rate, whereas energy absorption dynamics correlate with the strain rate. The amorphization efficiency can be accurately described by an \textit{effective temperature} model, while the energy absorption behavior adheres to the \textit{Herschel-Bulkley} constitutive relation. Furthermore, material properties play a pivotal role: The melting enthalpy of the crystalline phase emerges as a key determinant of the energy barrier within the effective temperature model, and the intrinsic amorphous state governs the rate of effective temperature increase. These parameters collectively enable a quantitative description of amorphization kinetics under oscillatory deformation, advancing mechanistic understanding and informing the design of novel brittle materials with tailored properties.
\section{Simulation Method}
\subsection{The initial sample preparation}

Three types of glass-forming alloy systems: CuZr, CuZr$_2$, Al$_3$Sm are used to investigate the transition of the crystal to the amorphous state induced by oscillatory deformation. \BSS{
The Al-Sm and Cu-Zr systems were selected for two principal reasons: (i) to ensure generalizability by investigating the amorphization mechanism across materials with distinct crystal symmetries, glass-forming abilities, and intrinsic properties; and (ii) to address the scarcity of suitable interatomic potentials, as these systems are well-characterized and exhibit good glass-forming ability.
}
The atomic interactions of all the systems are represented by the potential of the embedded atom method (EAM), for the Cu-Zr system\cite{Mendelev2019development} and the Al-Sm system\cite{Song2021molecular}, respectively.
The intermetallic single crystal structure is generated by the Atomsk package\cite{Hirel2015Atomsk}. 
For the CuZr system, the single crystal structure consists of 54,000 atoms with the B2 phase, and for the CuZr$_2$ system, the single crystal structure consists of 54,000 atoms with $C11_b$ phase, and for Al$_3$Sm system, the single crystal structure consists of 56,320 atoms with the phase $D0_{19}$. \BSS{Those crystal structure and pair distribution functions (PDFs) are shown in Figure\ref{fig:S0}.}
All systems were equilibrated in the constant pressure-constant temperature (NPT) ensemble at 300 K for 100 ps.

All simulations are carried out with the LAMMPS package\cite{Thompson2021LAMMPSA}, and atomic visualization is carried out with the OVITIO package\cite{Stukowski2009visualization}.

\subsection{The characteristic of atomic structure}

Two structural indicators are used in this work, one is orientation order, the crystal structure will display some orientation order, however the amorphous state is isotropic, we define order value for each atom to measure the orientation symmetry of atomic-level structure, the orientation order of one atom can be defined as\cite{PhysRevB.28.784}
\begin{equation}
   f_6 \equiv \frac{1}{N_c} \sum_{j \in N_c(i)} \Theta(S_6(i,j)-S_c)
\end{equation}
where $\Theta(x)$ is the step function, and $N_c(i)$ represents the number of neighbors of atom $i$, and $S_c$ is a threshold value of the order bond, in this work we use $S_c \equiv 0.7$, and $S_6(i,j)$ is orientations bound function, which can measure the order bond between two neighbor atoms, and particular sensitive with crystal phase studied in this work, namely, $B2$, $C11_b$ and $D0_{19}$ phase.

\begin{equation}
   S_6(i,j) \equiv \frac{\sum^6_{m=-6}{q_{6m}(i) \cdot q_{6m}^*(j)}}{\sqrt{\sum^6_{m=-6}{q_{6m}(i) \cdot q_{6m}^*(i)}}\sqrt{\sum^6_{m=-6}{q_{6m}(j) \cdot q_{6m}^*(j)}}}
\end{equation}

where $q_{6m}$ represents the standard bond-orientations parameter\cite{PhysRevB.28.784}, and $q_{6m}^{*}$ is the corresponding complex conjugate.

And the average value of $f_6$ in the whole system can be used to define the crystalline content $f_c$.  In a perfect single crystal state, $f_c=1$, whereas in an amorphous state, $f_c$ is low and approaches to zero\cite{Auer2004numerical,Russo2012microscopic}.

\subsection{The loading process of oscillatory deformation}
In this work, oscillatory deformations along three directions are used for mechanical amorphization.
The main direction is loading a sinusoidal strain as $\gamma_A sin(t/t_p)$, while the other directions used an opposite direction sinusoidal strain with half the strain amplitude, as $-\gamma_A/2 sin(t/t_p)$.
The primary coordinates axes and loading directions (tensile and compressive) can both be altered. Hence, for one configuration, there will be six deformation types for statistics.
For small strain amplitude deformation, the loading process is equivalent to pure shear deformation with conversation volume, and for large strain amplitude, the deformation can involve a density change, but the average process of volume change is not significant, and this kind of deformation is widely used in simulation to mimic mechanical alloying\cite{Rogachev2022MechanicalAI,Li2024stress} and ultrasonic loading\cite{shang2023influence,Li2023ultrafast}. 
%\BSS{The yield strain for above deformation protocol for  
%}
The temperature in all simulations is maintained at 300 K by the Nos\'e-Hoover thermostat\cite{nose1984unified}.
\subsection{\BSS{The melting enthalpy $\Delta H_m$}}
\BSS{
The melting enthalpy, $\Delta H_m$, which represents the enthalpy difference between the liquid and crystalline states at the melting temperature, was calculated for each material as follows. Single-crystal samples were heated from 300 K to 2500 K under an isothermal-isobaric (NPT) ensemble, employing a Nosé–Hoover thermostat\cite{nose1984unified} to control temperature. A constant heating rate of 0.1 K/ps was applied throughout the simulation. The transition from crystal to liquid was characterized by a discontinuous jump in enthalpy; the magnitude of this jump corresponds to $\Delta H_m$. For comparison, additional simulations were performed by quenching the system from 2500 K to 300 K at the same cooling rate of 0.1 K/ps to obtain the corresponding cooling curves.
}
\section{Results and Discussions}
\subsection{Loading process and Mechanical amorphization}
\begin{figure}[!]
    \centering
    \includegraphics[width=0.8\textwidth]{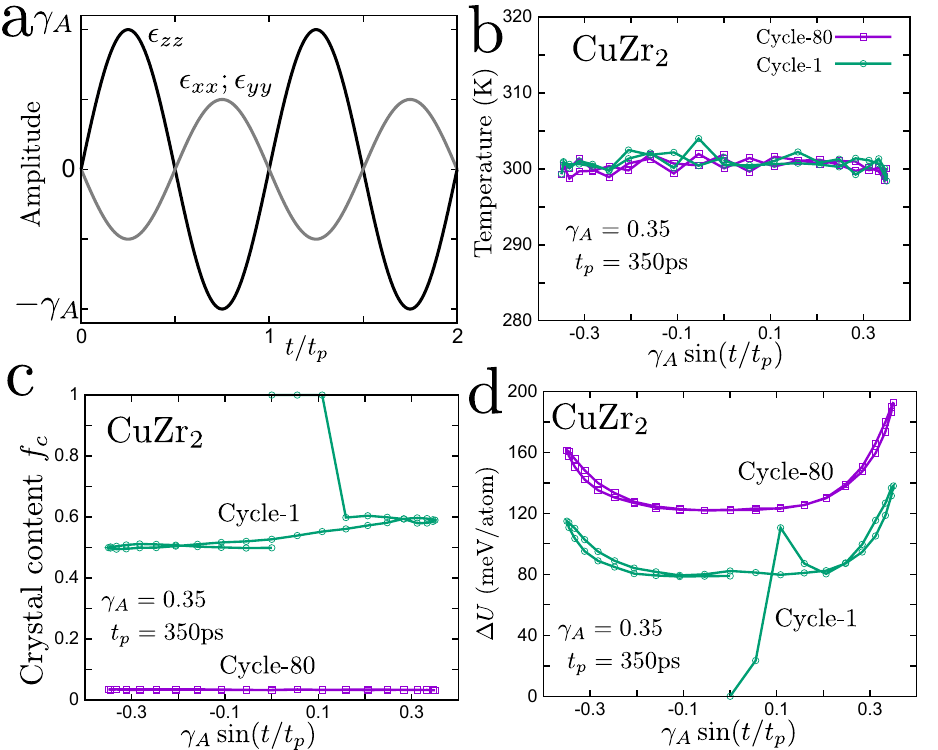}
\caption{\textbf{Oscillatory Deformation Loading Protocol}
    \textbf{a} Time-dependent loading strain profile: the main strain ($\epsilon_{zz}$) is twice the magnitude of the other strains ($\epsilon_{xx}$, $\epsilon_{yy}$) and is applied in the opposite direction.
    \textbf{b} Temperature fluctuations during the 1st and 80th cycles.
    \textbf{c} Crystal content $f_c$ evolution over the 1st and 80th cycles.
    \textbf{d} Changes in potential energy ($\Delta U$) across the 1st and 80th cycles.
}
    \label{fig:1}
\end{figure}

\begin{figure}[!]
    \centering
    \includegraphics[width=0.8\textwidth]{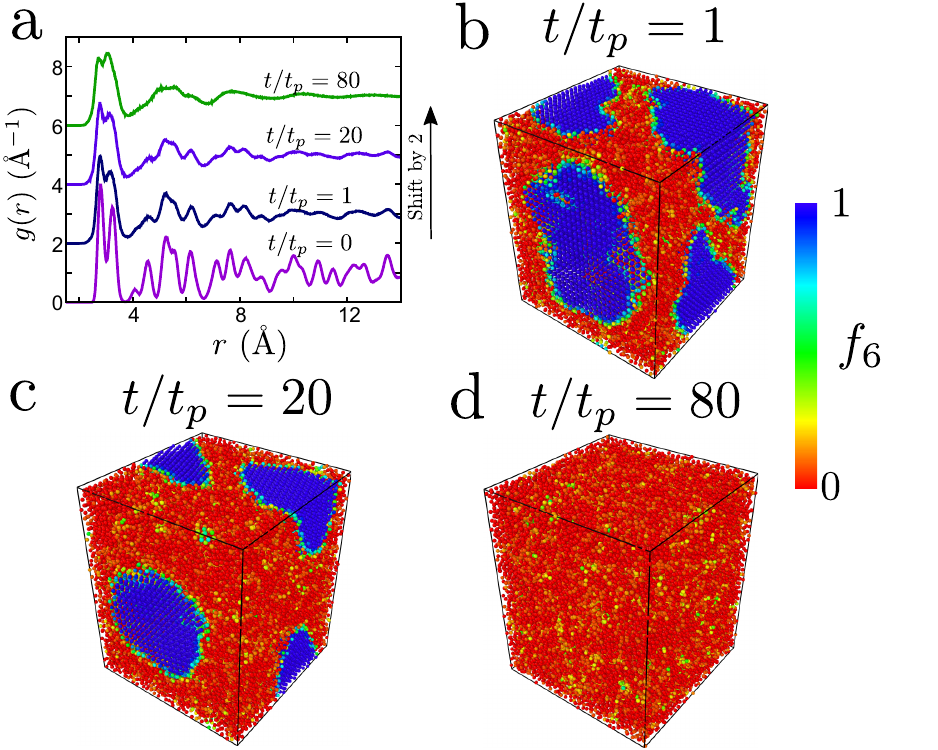}
\caption{\textbf{Evolution of Structure with Loading Cycle Number}. 
    \textbf{a} Evolution of the pair distribution function $g(r)$ with respect to $r$ for various cycles.
    \textbf{b--d} Spatial distribution of the atomic order parameter $f_6$ for the 1st cycle ($t/t_p=1$), 40th cycle ($t/t_p=40$), and 80th cycle ($t/t_p=80$), respectively. The material system is CuZr$_2$, with loading parameters $\gamma_A=0.35$ and $t_p=350$ ps.
}
    \label{fig:2}
\end{figure}

Figure \ref{fig:1} illustrates the loading protocol for oscillatory deformation within the CuZr$_2$ system. The primary strain applied is observed to be twice the magnitude of the other strains, which have opposite signs.
Upon comparing the first and 80th cycles, it is evident that the temperature remains consistently at the target value of 300 K (Figure \ref{fig:1}b). However, a substantial decrease in crystallization content ($f_c$) is observed, along with a significant increase in energy absorption (the potential energy difference $\Delta U$) between the initial crystalline state and the deformed system (Figure \ref{fig:1}c,d). These observations suggest that the amorphization process is primarily driven by mechanical deformation rather than thermal effects.

During the initial cycle, both the energy absorption ($\Delta U$) and the crystallization content ($f_c$) evolve in response to the applied load, as shown in Figure \ref{fig:1}c. In contrast, by the 80th cycle, $f_c$ shows little or no response to further external loading, with the sample stabilizing entirely in an amorphous state. This transition is corroborated by the data in Figure \ref{fig:2}, which shows the evolution of the pair distribution function $g(r)$ and the atomic order parameter $f_6$. As the number of mechanical cycles increases, the system undergoes a transformation from a crystal state to an amorphous state, accompanied by a reduction in the spatial extent of the crystalline regions as the external loading intensifies. In contrast to melting induced by thermal effects, this transition suggests that amorphization is driven by mechanical loading (Figure \ref{fig:2} b-d), with crystallization zones shrinking as the deformation progresses.
\subsection{The crystal to amorphous transition (CTA)}
   \begin{figure}[!htpb]
        \centering
        \includegraphics[width=0.8\textwidth]{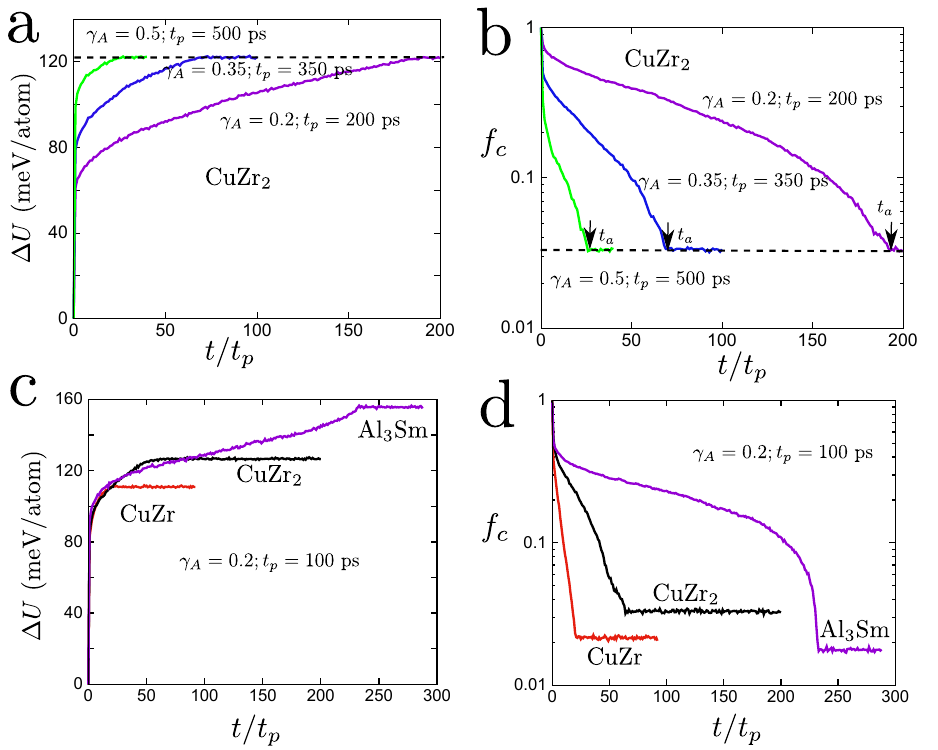}
        \caption{\textbf{Control Parameters of Oscillatory Deformation}
        \textbf{a, b} Evolution of energy absorption ($\Delta U$) and crystal content ($f_c$) over loading time for varying loading amplitudes ($\gamma_A$) and periods ($t_p$), respectively, in the CuZr$_2$ system. \BSS{
       Note the strain rate $\gamma/t_p$  is fixed at 0.001/ps.
        } The mechanical amorphization time ($t_a$) is indicated by arrows, and a dashed line is provided as a guide for the eye.
       \textbf{c, d} Evolution of energy absorption ($\Delta U_s$) and crystal content ($f_c$) over loading time for different material systems-CuZr, CuZr$_2$, Al$_3$Sm-under the same loading conditions: strain amplitude $\gamma_A = 0.2$ and loading period $t_p = 100$ ps.
        }
        \label{fig:3}
    \end{figure}

Figure \ref{fig:3} illustrates the change in energy absorption per cycle ($\Delta U$) and the crystallization content ($f_c$) as a function of the cycle number. After several cycles, both $\Delta U$ and $f_c$ reach a plateau, indicating a steady state within the material and signaling the completion of the crystal-to-amorphous transition (CTA). The time at which amorphization occurs, denoted $t_a$, can be identified by the onset of this plateau, as shown in Figure \ref{fig:3}b. This plateau represents a critical point in the mechanical amorphization process, where the material has absorbed a specific amount of energy and undergone a defined degree of deformation, leading to the formation of an amorphous structure.

Two key parameters define the mechanical amorphization process: one is steady-state energy absorption at the plateau ($\Delta U_s$), which quantifies the maximum energy that can be absorbed with a given load parameter. This energy corresponds to the work done on the material to drive the amorphization. The other key parameter is the amorphization time ($t_a$), which reflects the efficiency of the amorphization process. Specifically, $t_a$ represents the duration required for the material to undergo amorphization, serving as an indicator of the efficiency of the process under the given loading conditions.

In contrast to amorphization induced by high strain rate, such as shock loading or tensile deformation under uniaxial strain\cite{Ikeda1999StrainRI,Koh2006ShockinducedLA,Zhou2023Crystal}, which is characterized by a single parameter (strain rate, $\dot{\gamma}$), oscillatory deformation involves two key loading parameters: strain amplitude ($\gamma_A$) and loading period ($t_p$). The strain rate for oscillatory deformation ($\dot{\gamma}$) is proportional to the ratio ${\gamma_A}/{t_p}$. As shown in Figure \ref{fig:3}(a), for a series of values $\gamma_A$ and $t_p$ with the same ratio of ${\gamma_A}/{t_p}$, the potential energy on the plateau ($\Delta U_s$) remains constant; however, the amorphization time ($t_a$) exhibits variability.

This suggests that the steady-state energy ($\Delta U_s$) and the efficiency of the amorphization process ($t_a$) are governed by different parameters. Both $\Delta U_s$ and $t_a$ are sensitive to external loading conditions and material properties, as illustrated in Figures \ref{fig:3}(c) and (d).

\subsection{The effect of loading conditions and materials property on CTA}
\begin{figure}[!]
        \centering
        \includegraphics[width=0.8\textwidth]{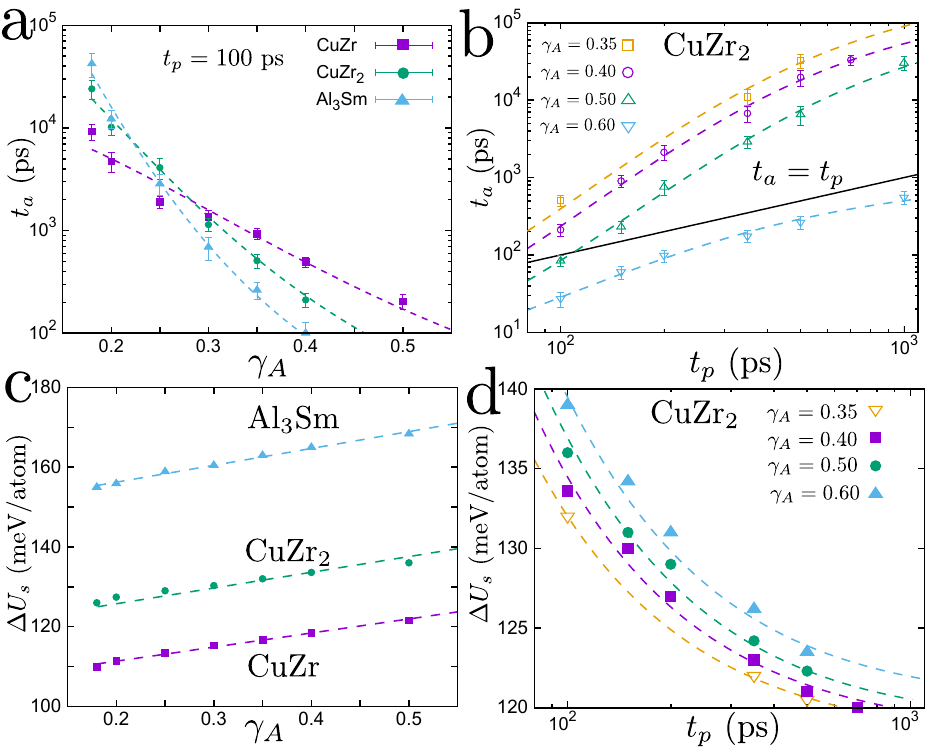}
\caption{\textbf{Dependence of Energy Absorption $\Delta U_s$ and Amorphization Time $t_a$ on Loading Parameters}. 
    \textbf{a} Mechanical amorphization time $t_a$ versus amplitude strain $\gamma_A$ for various systems, with a cycle period of $t_p=100$ ps.
    \textbf{b} $t_a$ versus period $t_p$ for the CuZr$_2$ system, with different amplitude strains $\gamma_A$. The solid line indicates the equivalence line $t_a = t_p$.
    \textbf{c} Potential energy $\Delta U_s$ versus amplitude strain $\gamma_A$ for $t_p=100$ ps.
    \textbf{d} $\Delta U_s$ versus period $t_p$ for different amplitude strains $\gamma_A$.
}
    \label{fig:4}
\end{figure}

Figure \ref{fig:4} illustrates the quantitative dependence of steady-state energy absorption ($\Delta U_s$) and amorphization time ($t_a$) on the loading conditions and the material systems.
For the amorphization time $t_a$, it decreases with increasing strain amplitude ($\gamma_A$) and increases with loading period ($t_p$) (Figure \ref{fig:4}a,b). In addition, a crossover behavior is observed for $t_a$ in different material systems (Figure \ref{fig:4}a). This behavior occurs when the relationship between the amorphization time and the strain amplitude is reversed for certain combinations of material and loading conditions. The sequence of amorphization efficiency for various materials can be modulated by strain amplitude, consistent with recent experimental findings in mechanical alloying\cite{Li2024stress}. %This suggests that the relationship between material properties and amorphization time is not universal and can be significantly influenced by external loading conditions.

However, for a given material system, the amorphization time $t_a$ consistently decreases with increasing strain amplitude ($\gamma_A$) and increases with increasing loading period ($t_p$), without exhibiting crossover behavior (Figure \ref{fig:4}b). This trend suggests that larger strain amplitudes facilitate a faster transition to the amorphous state, while longer loading periods allow for more gradual deformation, resulting in a longer transition time. Consequently, $t_a$ depends monotonically on both $\gamma_A$ and $t_p$.

Furthermore, the amorphization time $t_a$ exhibits a distinct behavior for different strain amplitudes. For larger strain amplitudes, $t_a$ is consistently shorter than the period time $t_p$, indicating that the transition does not complete a full periodic cycle. This suggests that the transition is predominantly determined by the strain amplitude ($\gamma_A$) rather than the loading period time ($t_p$). For example, when loading periods of up to 10 ns are tested (Figure \ref{fig:4} b and Figure \ref{fig:S1} in the Appendix), for $\gamma_A = 0.6$, the amorphization time $t_a$ is always shorter than $t_p$. In contrast, for smaller strain amplitudes, $t_a$ exceeds $t_p$, implying that the transition occurs over multiple periods and the amorphization process accumulates over the course of periodic loading. This behavior underscores the distinct roles of strain amplitude ($\gamma_A$) and loading period time ($t_p$), with $t_a$ being more sensitive to variations in $\gamma_A$ than to $t_p$.

The steady-state energy absorption $\Delta U_s$ increases with increasing strain amplitude ($\gamma_A$) and decreases with increasing loading period ($t_p$) (Figure \ref{fig:4}\BSS{c,}d). In contrast to $t_a$, $\Delta U_s$ exhibits a similar trend with respect to the strain amplitude between different material systems, without any crossover behavior. As the strain amplitude increases, $\Delta U_s$ tends to increase, indicating that larger strain amplitudes generally require more energy to induce amorphization. However, the magnitude of $\Delta U_s$ can vary significantly between materials, reflecting differences in their crystallographic structures and microstructural properties. This suggests that while the absorbed energy $\Delta U_s$ tends to be higher for larger strain amplitudes, the final energy state can differ significantly between materials, potentially affecting their overall mechanical properties and behavior.

Furthermore, $\Delta U_s$ maintains a consistent trend between different strain amplitudes (Figure \ref{fig:4}d), indicating that the relationship between the loading conditions and the energy absorbed is consistent regardless of the strain amplitude. This suggests that the energy required to induce amorphization is primarily dependent on the deformation rate rather than the absolute magnitude of the strain amplitude.

\subsection{The control parameters of CTA: work rate and strain rate}
\begin{figure}[!]
    \centering
    \includegraphics[width=0.8\textwidth]{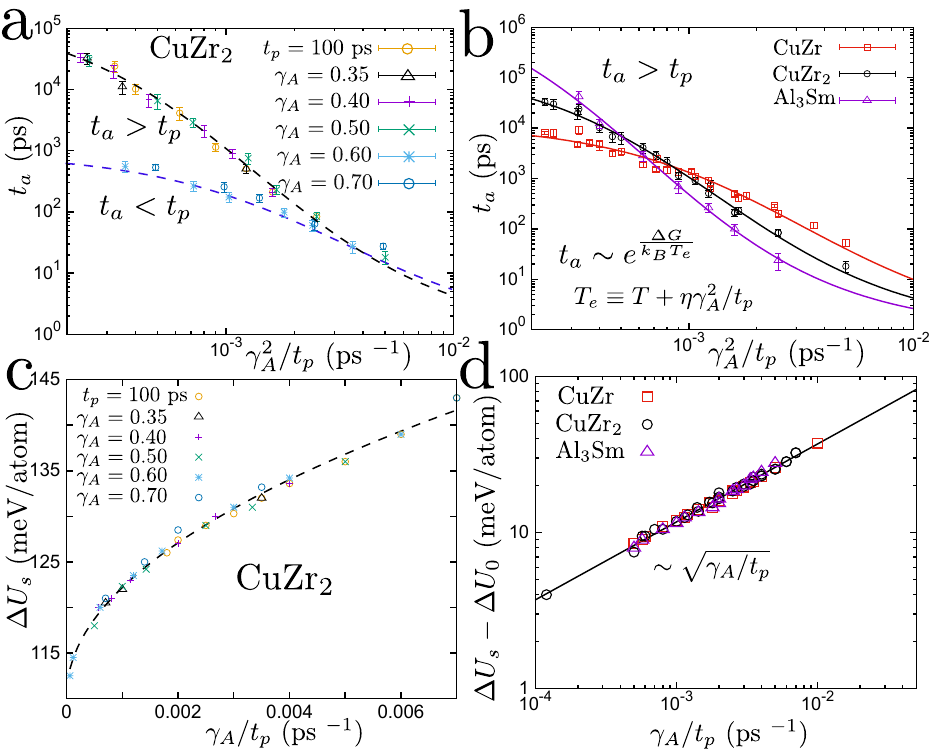}
    \caption{
    \textbf{Dependence of Mechanical Amorphization Time $t_a$ on Work Rate $\gamma_A^2/t_p$ and Energy Absorption $\Delta U$ on Strain Rate $\gamma_A/t_p$.}
    \textbf{a} $t_a$ versus work rate $\gamma_A^2/t_p$ in the CuZr$_2$ system.
    \textbf{b} $t_a$ versus $\gamma_A^2/t_p$ for the CuZr, CuZr$_2$, Al$_3$Sm systems, where $t_a > t_p$. The solid line is a fit to the effective temperature model $t_a = t_0 e^{\Delta G/k_B T_e}$, with $t_0 \equiv 1$ ps.
    \textbf{c} $\Delta U_s$ versus strain rate $\gamma_A/t_p$ in the CuZr$_2$ system.
    \textbf{d} $\Delta U_s-\Delta U_0$ versus $\gamma_A/t_p$ for the CuZr, CuZr$_2$, Al$_3$Sm systems. The solid line is a fit to the Herschel-Bulkley relation $\Delta U_s = \Delta U_0 + B t_{0}^{1/2} (\gamma_A/t_p)^{1/2}$.
    Dashed lines in \textbf{a} and \textbf{c} are guided for the eye.
    }
   \label{fig:5}
\end{figure}

Figure \ref{fig:5} shows the mechanical amorphization time $t_a$ and steady-state energy absorption $\Delta U_s$ as functions of the work rate (${\gamma_A^2}/{t_p}$) and the strain rate (${\gamma_A}/{t_p}$), respectively. Interestingly, all the data from Figure \ref{fig:4} collapse onto a master curve, each represented by different variables. For the amorphization time $t_a$, it is a function of the work rate, while for the steady-state energy $\Delta U_s$, it depends on the strain rate. The different functional dependencies highlight different behaviors and offer a practical approach to independently tune the efficiency and energy state of the amorphization process.

Moreover, the relationship between $t_a$ and the work rate can be further categorized by its dependence on the loading period $t_p$, with this behavior being controlled by the strain amplitude $\gamma_A$. For sufficiently large $\gamma_A$, $t_a$ is always less than $t_p$, regardless of the work rate. This indicates that the strain amplitude $\gamma_A$ is the dominant parameter, rather than $t_p$, in determining the completion of amorphization within a single cycle. This effect could be attributed to the significant changes in system volume induced by larger deformations. In this regime, the behavior is analogous to uniform loading at high strain rates\cite{Zhou2023Crystal}, although the key parameter is the work rate rather than the strain rate.

In contrast, for small $\gamma_A$, the amorphization time $t_a$ exceeds the loading period $t_p$, with the amorphization process extending over multiple cycles. This behavior is consistent with the universal master curve. For practical experimental applications, we focus on the regime where $t_a > t_p$, which aligns with processes such as ultrasonic fabrication\cite{Li2023ultrafast} or ball grinding\cite{Li2024stress}. As the crystalline zone shrinks and the amorphous zone grows under mechanical loading (see Figure \ref{fig:2}b-d), this process becomes analogous to a melting phenomenon, even when the temperature is maintained at room temperature (Figure \ref{fig:1} b), the effective temperature can increase due to external loading\cite{Cugliandolo2011effective, PhysRevLett.104.205701}.
The effective temperature model, which has been used to describe crystallization in supercooled liquids under external loading\cite{PhysRevE.87.062307,shang2023influence}, can be adapted to describe mechanical amorphization. Unlike crystallization under external loading, where the strain rate or stress is typically the controlling factor\cite{PhysRevE.87.062307,shang2023influence}, here we find that the effective temperature is governed by the work rate in mechanical amorphization induced by oscillatory deformation.

The master curve for $t_a > t_p$ can be well described by an effective temperature model, given by: 
\begin{equation} 
t_a = t_0 e^{\frac{\Delta G}{k_B T_\text{e}}} 
\label{eqn:1} 
\end{equation} 
Where $t_0 \equiv 1$ ps for dimensional balance, $\Delta G$ is the free energy barrier for the crystal-to-amorphous transition in the absence of loading, and $T_\text{e} \equiv T + \eta t_0 \left(\frac{\gamma_A^2}{t_p}\right)$ is the effective temperature. The effective temperature consists of two contributions: one from the system's temperature and the other induced by the external loading, which is assumed to be linearly dependent on the work rate $\gamma_A^2/t_p$. \BSS{Note this linear dependence between the effective temperature and the work rate is merely an approximation derived from a Taylor expansion}. Here $\eta$ is a material-dependent parameter that characterizes the rate of change in the effective temperature with the increasing work rate.

Regarding the amorphous energy $\Delta U_s$, all data from different strain rates and material systems collapse into a single master curve (Figure \ref{fig:5}d), suggesting that $\Delta U_s$ is only determined by the strain rate and the quiescent amorphous energy ($\Delta U_0$) of the system.  This master curve can be described using the \textit{Herschel-Bulkley} constitutive relation, which typically describes the relationship between flow stress and strain rate\cite{RevModPhys.89.035005}. Similar phenomena have been observed in simple shear loading and other amorphous solid systems\cite{PhysRevLett.103.065501,Nicolas2016}, indicating a universal dependence between the shear rate and the state of the system. In this study, we find that this relationship also holds for $\Delta U_s$ and strain rate $\gamma_A/t_p$, which can be described by the following equation: 
\begin{equation} 
\Delta U_s = \Delta U_0 + B \sqrt{t_0} \sqrt{\frac{\gamma_A}{t_p}} 
\label{eqn:2} 
\end{equation} 
where $\Delta U_0$ is the quiescent amorphous energy, determined by material properties, $t_0 \equiv 1 $ ps for the dimensional balance and $B$ is a fitting constant that is independent of material properties and loading protocol. The value of $B$ is $370.5 \pm 1.4$ meV/atom.

\begin{figure}[!h]
        \centering
        \includegraphics[width=0.8\textwidth]{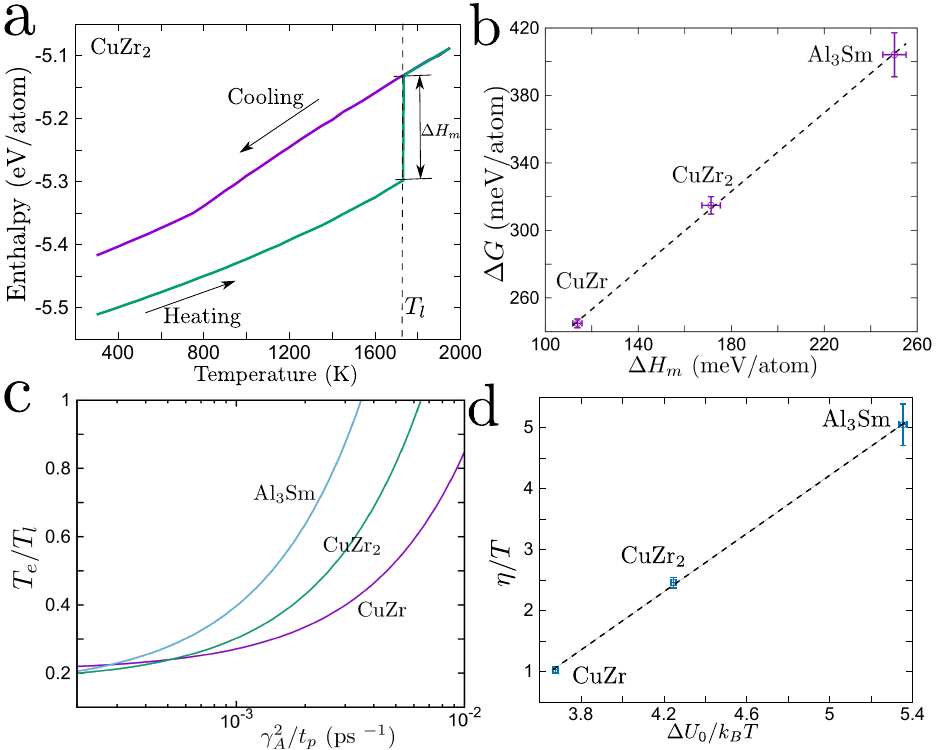}
        \caption{
        \textbf{The connection between thermodynamic property and mechanical amorphization}
        \textbf{a.} The temperature-enthalpy map in heating-cooling process in $CuZr_2$ system. The single crystal $CuZr_2$ was heating from 300 K to 2000 K with heating rate $10^{12}$ K/s, and then cooling from 2000 K to 300 K with cooling rate $10^{12}$ K/s. The enthalpy change during heating process is melting enthalpy $\Delta H_{m}$. And the corresponding temperature is denoted as the melting temperature $T_l$.
        \textbf{b.} the correlation between melting enthalpy $\Delta H_m$ and free energy barrier $\Delta G$ fitted by effective temperature model for CuZr, CuZr$_2$, Al$_3$Sm system.
        \textbf{c.} the reduced effective temperature $T_e/T_l$ evolves with work rate $\gamma_A^2/t_p$ for three systems.
        \textbf{d.} The correlation between quiescent potential energy $\Delta U_0/k_BT$ and mechanical temperature $\eta/T$ for three systems.
        The dash lines in \text{b.} and \text{d.} are guided for the eye.
        }
        \label{fig:6}
\end{figure}

\BSS{
The measured mechanical response is well-captured by a Herschel-Bulkley constitutive relation. However, while this model provides an excellent empirical fit, the physical origin of the observed strain-rate scaling exponent of n = 1/2 warrants further discussion. This specific value may not be arbitrary but could emerge from the underlying micromechanics of the amorphous phase, particularly the dynamics of Shear Transformation Zones (STZs)\cite{PhysRevE.57.7192}. We hypothesize that the exponent of 1/2 is linked to a fundamental scaling between the applied strain rate and the activation energy for STZ operation. In amorphous solids, plastic flow is often governed by a stress-dependent energy barrier, $\Delta E \sim 1/\tau$, where $\tau$ is the shear stress\cite{Falk2011deformation}. If the effective stress governing STZ activation scales with the square root of the strain rate ($\tau_{\text{eff}} \sim \dot{\gamma}^{1/2}$), it would naturally yield a Herschel-Bulkley exponent of n = 1/2. This scaling could arise from a mean-field interaction between STZs or from a specific form of cooperative, diffusive-like motion within the shear bands. This proposed connection remains speculative and highlights a critical area for future investigation.
}

\BSS{Furthermore, our results demonstrate that the steady-state energy absorption, $\Delta U_s$, varies significantly across different material systems. This variation is primarily attributed to the intrinsic material property $\Delta U_s$, the energy offset between crystalline and amorphous states, which correlates with the melting enthalpy, $\Delta H_m$  (see Fig.\ref{fig:S4}). This suggests that the fundamental thermodynamic stability of the crystal relative to the glass plays a key role in determining how much energy a material can absorb during mechanical amorphization.

However, a direct correlation between traditional microstructural order parameters, such as bond orientation order, and $\Delta U_s$ remains elusive. For instance, the crystal fraction ($f_c$) exhibits trends divergent from $\Delta U_s$ and is itself invariant to loading parameters in the amorphous state  (Figure \ref{fig:3}c,d). The precise microstructural features that govern energy absorption thus present a compelling open question for future research, potentially requiring more sophisticated descriptors of amorphous structure.}

So far, for external cyclic loading, the efficiency of mechanical amorphization is determined by the work rate, while the amorphous energy is controlled by the strain rate. In contrast, the free energy barrier $\Delta G$, the mechanical temperature unit $\eta$ (as described in Equation \ref{eqn:1}) and the quiescent amorphous energy $\Delta U_0$ are solely dependent on the topological structure of the energy landscape, regardless of the loading conditions.

To better understand the physical significance of these parameters, we compare mechanical amorphization with the heating-induced melting of a crystalline structure. There are generally two pathways for the crystal-to-amorphous transition: one is induced by melting and the other by mechanical deformation. As shown in Figure \ref{fig:6}(a), during the reheating process, the crystal-to-liquid transition is marked by a jump in enthalpy, known as the melting enthalpy ($H_m$). This melting enthalpy is largely insensitive to the heating rate and reflects the free energy difference between the liquid and crystalline states. In particular, the melting enthalpy ($H_m$) exhibits a linear relationship with the energy barrier $\Delta G$ (Figure \ref{fig:6}(b)). This finding suggests an intrinsic connection between melting and mechanical amorphization transitions and provides a practical experimental method for measuring the energy barrier of mechanical amorphization.

As expressed in Equation \ref{eqn:1}, the efficiency of mechanical amorphization is determined by both the effective temperature caused by external loading ($T_e$) and the energy barrier ($\Delta G$). In Figure \ref{fig:6}(c), we present the rescaled effective temperature, $T_e/T_l$, where $T_l$ is the melting temperature, as a function of the work rate. For low work rates, the effective temperature is similar across different systems and is primarily controlled by the environmental temperature. However, at high work rates, the three systems exhibit different rate of increase, which are mainly governed by the mechanical temperature unit ($\eta$).

Moreover, $\eta$ shows a linear relationship with the quiescent amorphous energy $\Delta U_0$ and a positive correlation with the energy barrier $\Delta G$ (Figures \ref{fig:6}(d) and \ref{fig:S3}). This suggests that the mechanical temperature unit is intricately determined by the properties of the material, independent of the energy state.

At low work rates, the efficiency of mechanical amorphization is mainly controlled by the energy barrier $\Delta G$. Since $\Delta G$ is approximately proportional to the melting enthalpy $\Delta H_m$, glass formation systems with smaller $\Delta H_m$ exhibit shorter mechanical amorphization times. At higher work rates, the effective temperature increases significantly, dominating the efficiency of the process. In this regime, $\eta \sim \Delta U_0$, and materials with larger quiescent amorphous energies $\Delta U_0$ tend to have shorter mechanical amorphization times. Interestingly, there is also a positive correlation between $\Delta H_m$ and $\Delta U_0$ (as well as between $\Delta G$ and $\eta$, see Figures \ref{fig:S3} and \ref{fig:S4}).

This leads to the conclusion that the crossover behavior observed in the mechanical amorphization process between different systems (Figures \ref{fig:4}(a) and \ref{fig:5}(b)) can be primarily determined by the melting enthalpy. This finding is consistent with recent experimental work\cite{Li2024stress}, which links the ability to form glass with the mechanical amorphization ability.

\begin{figure}[!htpb]
    \centering
    \includegraphics[width=0.8\linewidth]{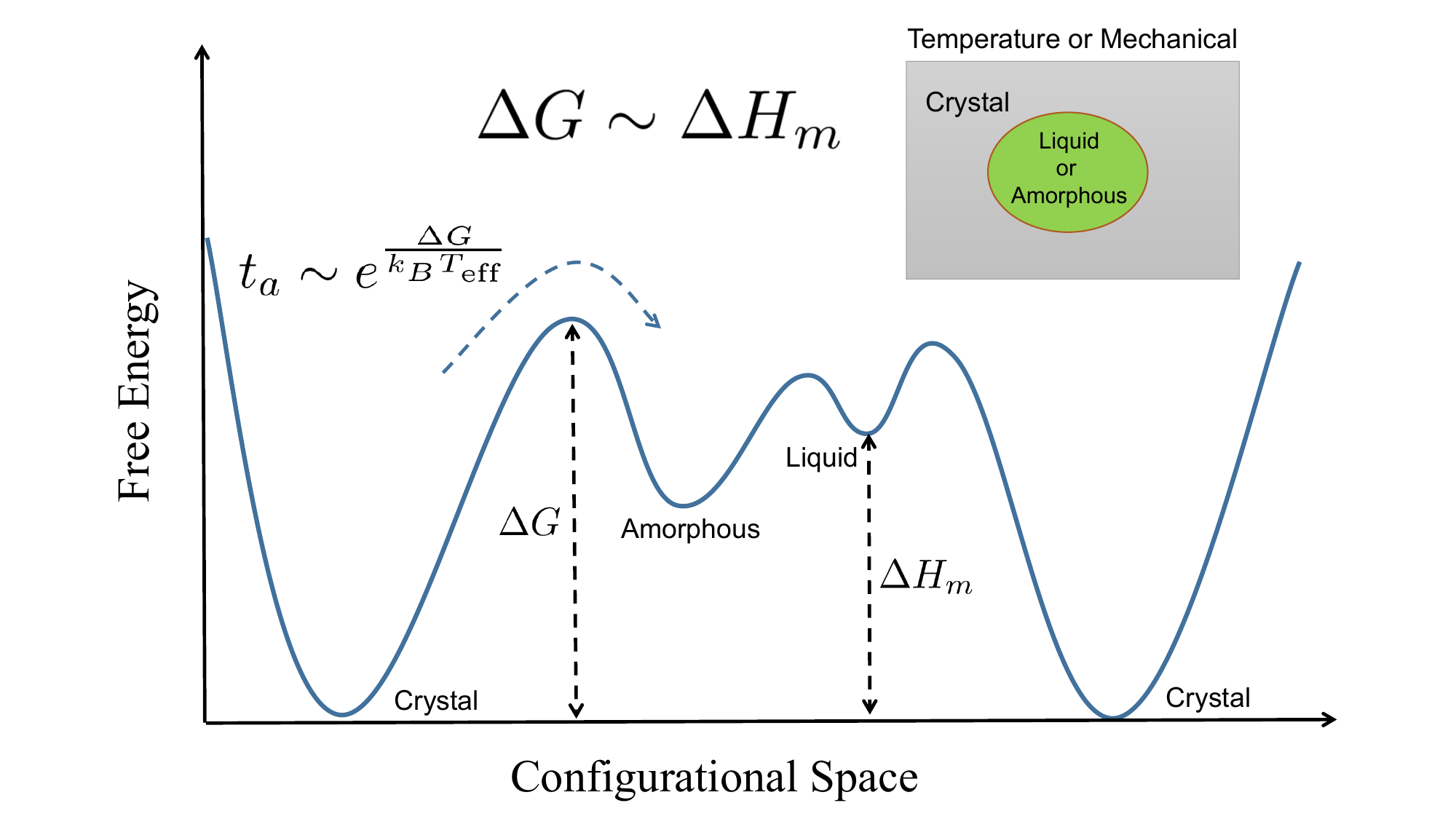}
    \caption{\BSS{
     A schematic of free energy landscape for amorphous solid, liquid and crystal. In the panel, a disorder phase (liquid or amorphous) grows in the crystal phase.
    }
    }
    \label{fig:7}
\end{figure}
\BSS{
The relationship between melting enthalpy, $\Delta H_m$, and the mechanical amorphization energy barrier, $\Delta G$, is illustrated in Figure \ref{fig:7}. In good glass-forming materials, both mechanical amorphization and thermal melting can be viewed as processes involving the nucleation of a disordered phase within an ordered crystalline matrix. The observed positive correlation between $\Delta G$ and $\Delta H_m$ can be rationalized within the framework of classical nucleation theory (CNT), wherein the free energy barrier is governed by two competing contributions: the volumetric thermodynamic driving force (which promotes the phase transformation and is related to $\Delta H_m$) and the interfacial energy penalty (which opposes it).

The interfacial energy term plays a dominant role in this correlation. In a highly stable crystal—characterized by a large $\Delta H_m$, forming a disordered interface requires breaking strong, ordered atomic bonds, resulting in a high energy cost. Thus, materials with greater $\Delta H_m$ exhibit higher interfacial energies. The positive dependence of $\Delta G$ on $\Delta H_m$ (see Figure \ref{fig:6}) indicates that the increased interfacial energy penalty in such stable crystals outweighs the effect of the larger thermodynamic driving force. This highlights that kinetic stability against amorphization is significantly influenced by the difficulty of interface formation.
}

\section{Conclusion}

In this work, we systematically investigated the energy absorption and efficiency of mechanical amorphization induced by oscillatory deformation in three typical glass-forming systems. Based on the detailed analysis presented, several key conclusions can be drawn regarding the mechanical amorphization process and its dependence on external loading conditions and topological energy landscape of materials.

First, the efficiency of mechanical amorphization is primarily governed by the work rate, which can be well depicted by the effective temperature model, while the energy absorption is determined by the strain rate, which can be depicted by the \textit{Herschel-Bulkley} constitutive relation. Moreover, the effective temperature model, which accounts for both the system temperature and the contributions from external loading, successfully describes the efficiency of mechanical amorphization, and the key parameters are determined by the topological energy landscape of materials: the free energy barrier between amorphous state and the quiescent amorphous energy. The model reveals that, at low work rates, efficiency is primarily controlled by the free energy barrier, whereas at high work rates, the effective temperature plays a dominant role. This dual dependence underscores the complex interaction between the properties of the material and external loading conditions to determine the rate of amorphization.

A notable insight from this study is the analogy between mechanical amorphization and heating-induced melting of crystalline structures. Both processes involve a transition from a crystalline state to an amorphous or liquid state, with the melting enthalpy reflecting the free energy difference between the two states. The study reveals a direct relationship between the melting enthalpy and the energy barrier for mechanical amorphization , suggesting that the two transitions are intrinsically linked. This relationship provides a practical experimental method to measure the energy barrier for mechanical amorphization and highlights a universal trend that can be applied across various material systems.

The study also identified a crossover behavior in the mechanical amorphization efficiency across different material systems, which is governed by the melting enthalpy. This finding is consistent with recent experimental work linking glass-forming ability to mechanical amorphization capacity, suggesting that materials with lower $H_m$ exhibit shorter mechanical amorphization times. This insight is crucial for tailoring materials for specific applications, particularly in processes such as ultra-fast fabrication and ball milling.

In conclusion, this work provides a comprehensive framework for understanding the mechanical amorphization process. By elucidating the interaction between material properties, external loading conditions, and the energy landscape of the transition, we established a clearer pathway for controlling and optimizing the amorphization process in materials science. The findings not only advance our understanding of mechanical deformation-induced phase transitions but also offer practical insights for designing materials with tailored amorphization properties, with potential applications in ultra-fast fabrication, ball milling, and other mechanical processing techniques.

\section{Acknowledgments}
The authors thank Prof. Haiyang Bai, Prof. Jean-Louis Barrat, and Prof. Mo Li for insightful discussion. 
\section{Funding} 
 This work is supported by the Pearl River Talent Recruitment Program (Grant No.2021QN02C04), the NSF of China (Grant Nos.12474187, 52130108, 52201176) and Young Talent Support Project of Guangzhou Association for Science and Technology (Grant No.QT2024–041).
\section{Author contributions}
B.S.S. and X.X.L contributed equally to this work. B.S.S, P.F.G and W.H.W. conceived and supervised this work; B.S.S designed and conducted all the simulations with assistance from X.X.L; B.S.S., X.X.L., and P.F.G. wrote the paper with input and comments from all authors.
%\bibliographystyle{unsrt} 
%\bibliography{ref.bib}

\newpage
 \section*{Appendix}
\renewcommand{\thefigure}{A\arabic{figure}}
\renewcommand{\theequation}{A\arabic{equation}}
\setcounter{figure}{0}
\setcounter{equation}{0}
\begin{figure}[!htpb]
    \centering
    \includegraphics[width=0.8\linewidth]{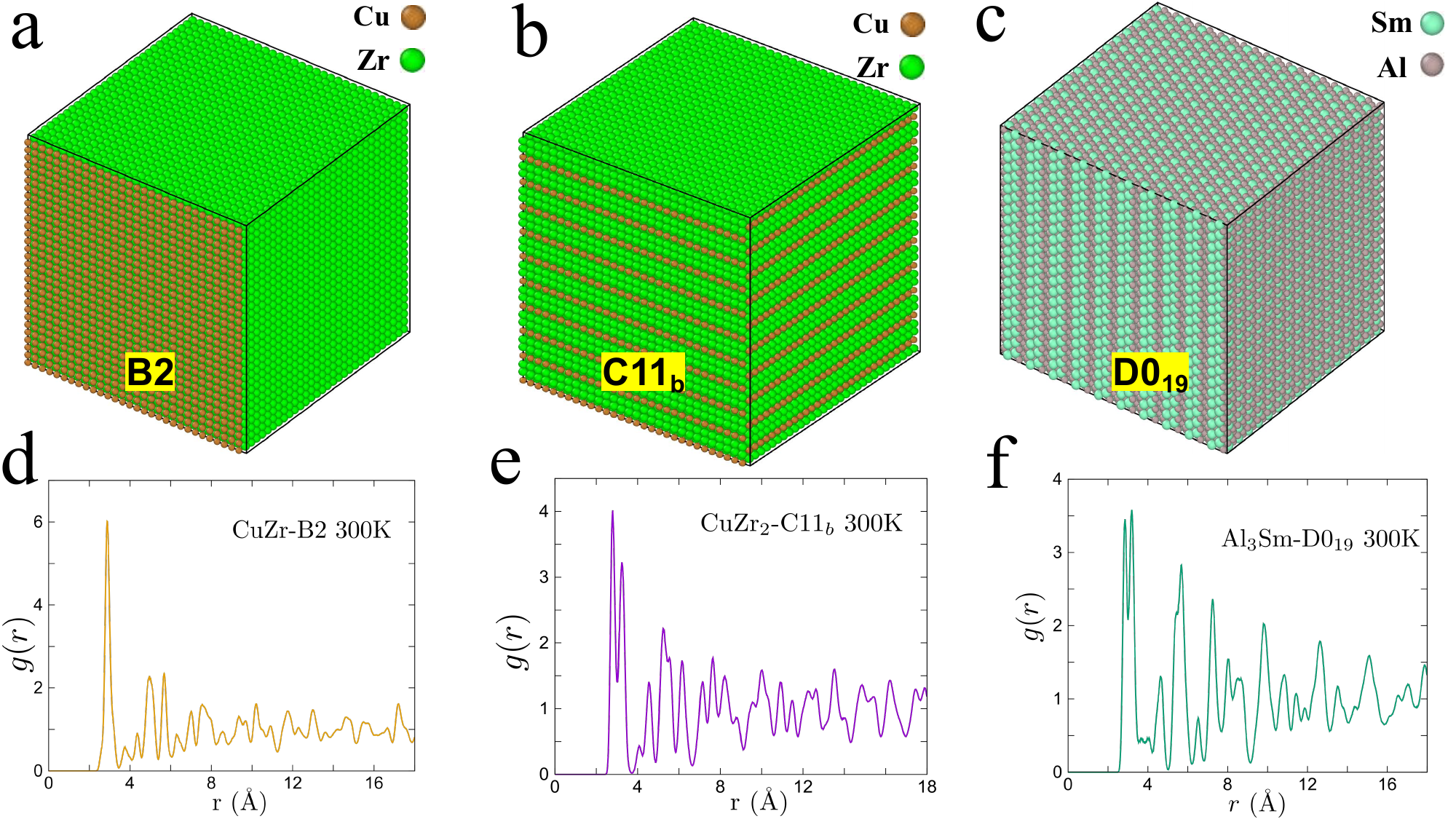}
    \caption{\BSS{
     Crystal structures of the three model systems: \textbf{a} CuZr, \textbf{b} CuZr$_2$, and \textbf{c} Al$_3$Sm. The corresponding radial distribution functions (RDFs) calculated at 300 K are presented in panels \textbf{d}, \textbf{e}, and \textbf{f}, respectively.
    }
    }
    \label{fig:S0}
\end{figure}
\begin{figure}[!htpb]
    \centering
    \includegraphics[width=0.5\linewidth]{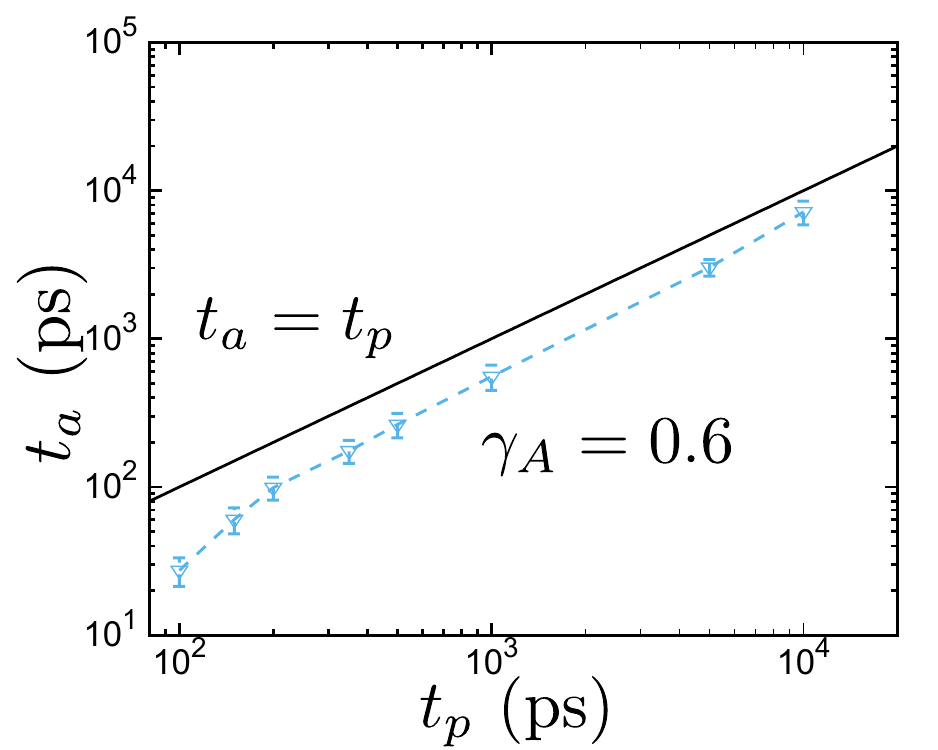}
    \caption{$t_a$ versus $t_p$ for CuZr$_2$ system with $\gamma_A=0.6$, the solid line illustrates $t_a=t_p$}
    \label{fig:S1}
\end{figure}
\begin{figure}[!htpb]
    \centering
    \includegraphics[width=0.5\linewidth]{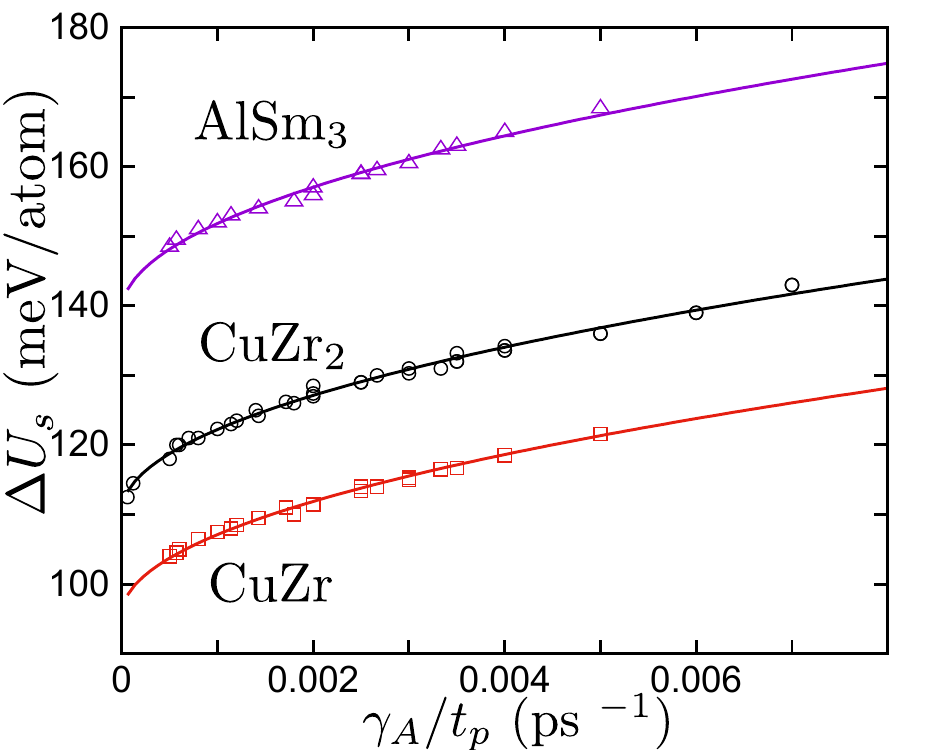}
    \caption{$\Delta U_s$ versus $\gamma_A/t_p$ for all the systems. The solid line is a fit to the Herschel-Bulkley relation $\Delta U_s = \Delta U_0 + B (\gamma_A/t_p)^{1/2}$}
    \label{fig:S2}
\end{figure}
\begin{figure}[!htpb]
    \centering
    \includegraphics[width=0.5\linewidth]{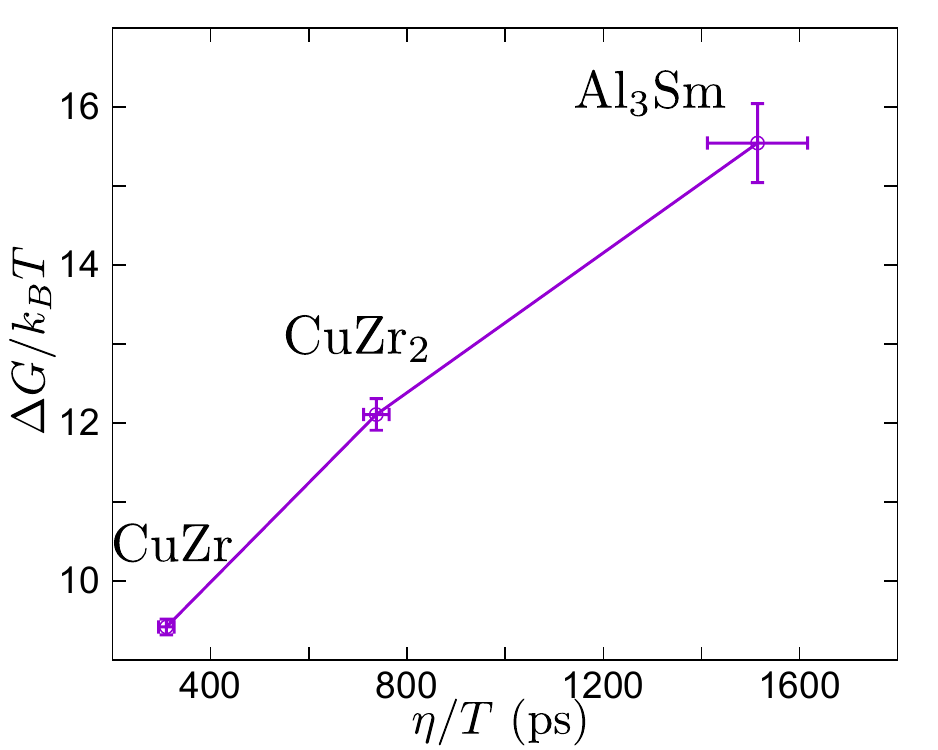}
    \caption{The correlation between energy barrier $\Delta G$ and mechanical temperature unit $\eta$ for all three systems.}
    \label{fig:S3}
\end{figure}
\begin{figure}[!htpb]
    \centering
    \includegraphics[width=0.5\linewidth]{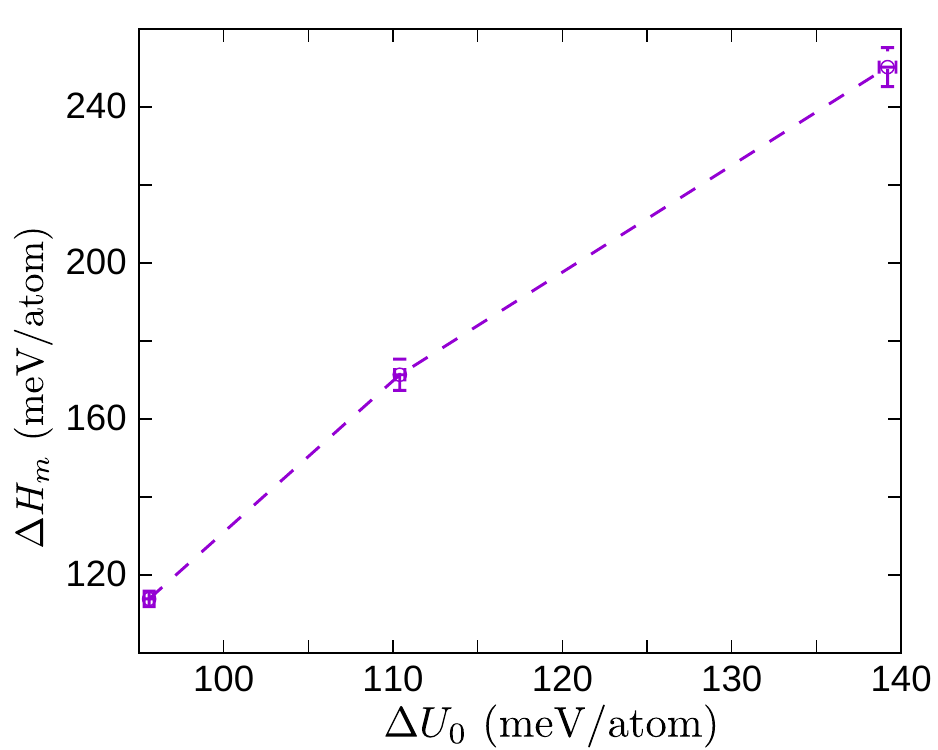}
    \caption{The correlation between quiescent amorphous energy $\Delta U_0$ and melting enthalpy $\Delta H_m$ for all three systems}
    \label{fig:S4}
\end{figure}
%\section{Graphical Abstract}
%\renewcommand{\thefigure}{\arabic{figure}}
%\setcounter{figure}{0}
%\begin{figure}
%    \centering
%    \includegraphics[width=1.0\linewidth]{figure/graphical-abstract-04-15.pdf}
%    \caption{}
    %\label{fig:enter-label}
%    \caption{\textbf{Graphical Abstract}: \textbf{(a)} the schematic of amorphization efficiency and absorption energy on the two-state potential landscape. \textbf{(b)} amorphization efficiency controlled by work rate. \textbf{(c)} absorption energy controlled by strain rate.}
%\end{figure}
\end{document}